\documentclass[
 reprint,
 amssymb, amsmath,
 aip,jcp                                                                                                                                                
]{revtex4-1}

\usepackage[english]{babel}

\usepackage[twocolumn,a4paper,top=1.27cm,bottom=1.27cm,left=1.27cm,right=1.27cm,marginparwidth=1.75cm]{geometry}

\usepackage{amsmath,amssymb,bm}
\usepackage{comment}
\usepackage{mathtools} 
\usepackage{graphicx}
\usepackage{xcolor}
\usepackage{physics}
\usepackage[colorlinks=true, allcolors=blue]{hyperref}
\usepackage{wrapfig}
\usepackage{lipsum,booktabs}
\usepackage{bm}
\usepackage{rotating}
\usepackage{multirow}
\usepackage{float}
\usepackage{adjustbox}

\newcommand{\basis}[0]{\mathcal{B}}
\newcommand{\elemm}[3]{\bra{#1} #2 \ket{#3}}

\newcommand{\vnn}[0]{V_{\text{NN}}}
\newcommand{\hkin}[0]{\hat{T}}
\newcommand{\hqmmm}[0]{\hat{H}_{\text{tot}}}

\newcommand{\hvne}[0]{\hat{V}_\text{Ne}}
\newcommand{\vqmmm}[0]{\hat{V}_{\text{C}}}

\newcommand{\hwee}[0]{\hat{W}_\text{ee}}
\newcommand{\brnucl}[1]{{\bf R}_{#1}}
\newcommand{\br}[1]{{\boldsymbol r}_{#1}}
\newcommand{\brr}[0]{{\boldsymbol r}}

\begin{document}

\title{Coupling Molecular Density Functional Theory with Converged Selected Configuration Interaction Methods to Study Excited states in Aqueous Solution}

\author{Maxime Labat}
\affiliation{Sorbonne Universit\'e, CNRS, Physico-Chimie des \'electrolytes et Nanosyst\`emes
Interfaciaux, PHENIX, F-75005 Paris, France}

\author{Emmanuel Giner$^*$}
\email{emmanuel.giner@lct.jussieu.fr}
\affiliation{Sorbonne Universit\'e, CNRS, Laboratoire de Chimie Th\'eorique, Sorbonne Universit\'e, F-75005 Paris, France}

\author{Guillaume Jeanmairet$^*$}
\email{guillaume.jeanmairet@sorbonne-universite.fr}
\affiliation{Sorbonne Universit\'e, CNRS, Physico-Chimie des \'electrolytes et Nanosyst\`emes
Interfaciaux, PHENIX, F-75005 Paris, France}
\affiliation{R\'eseau sur le Stockage \'electrochimique de l'\'energie (RS2E), FR CNRS
3459, 80039 Amiens Cedex, France}

\begin{abstract}
This paper presents the first implementation of a coupling between advanced wave function theories and molecular density functional theory (MDFT). This method enables the modeling of solvent effect into quantum mechanical (QM) calculations by incorporating an electrostatic potential generated by solvent charges into the electronic Hamiltonian. Solvent charges are deduced from the spatially and angularly dependent solvent particle density. Such density is obtained through the minimization of the functional associated to the molecular mechanics (MM) Hamiltonian describing the interaction between the fluid particles. The introduced QM/MDFT framework belongs to QM/MM family of methods but its originality lies in the use of MDFT as the MM solver, offering two main advantages. Firstly, its functional formulation makes it competitive with respect to sampling-based molecular mechanics. Secondly, it preserves a  molecular-level description  lost in macroscopic continuum approaches. Excited states properties of  water and formaldehyde molecules solvated into water have been computed at the selected configuration interaction (SCI) level. Excitation energies and dipole moment have been compared with experimental data and previous theoretical work. A key finding is that using the Hartree-Fock method to describe the solute allows for predicting the solvent charge around the ground-state with sufficient precision for the subsequent SCI calculations of excited-states. This significantly reduces the computational cost of the described procedure, paving the way for the study of more complex molecules.


\end{abstract}

\maketitle

\section{1. Introduction\label{sec:Intro}}
{Most of the chemical properties, including  reactivity and spectroscopy, are governed by electronic properties. This is the reason why electronic structure calculations play  a central role in theoretical and computational chemistry. Energy differences should be computed with a precision of typically 0.05 eV (approximately $1\  \text{kcal.mol}^{-1}$) for reactivity and in the range ${10^{-4}\text{ to }10^{-1}\ \text{eV}}$ depending on the targeted spectroscopic property. 
To be able to compare with actual experimental results, one needs to account for the environment in electronic structure calculations, as it can significantly influence the properties of the studied molecules.
The most ubiquitous effect to consider is likely the influence of the solvent, given that the majority of chemical processes occur in solution rather than in the gas phase. In organic chemistry, the solvent has essentially a two-fold role. It acts as a medium that brings the reactant species together, but it also subtly influences the reaction path. This aspect has been acknowledged and exploited by chemists since the early 30s\cite{Hughes-JCS-35}. In spectroscopy,  the solvent impacts the excitation energies of the analyte modifying its spectroscopic signature, the most spectacular example being solvatochromism\cite{machado_easy_2001}. }

Regarding reactivity, the solvent can  modify the reaction profile by  stabilizing or destabilizing the reactants and products. Moreover, it can also be directly involved in the reaction mechanism, for instance by forming chemical bonds with the reactants. This is also true in spectroscopy where  the solvent can actively contribute to the excitation process (\textit{i.e.} its molecular orbitals are explicitly involved in the electronic excited and/or ground states)  or play a passive role when the excitation process is located on the solute.

The most straightforward way to account for solvent effect is to explicitly include solvent molecules into the QM calculations. However, this  considerably increases the computational cost which scales as a polynomial of the number of electrons. Moreover, neglecting temperature in solution is a much cruder assumption than in vacuum (gas phase). Indeed, a solution is a liquid whose constituting molecules are in constant motion. 
This implies that one should no longer perform  a unique quantum mechanics (QM) calculation  for a given set of coordinates but rather sample the solute and solvent configurations until a statistically converged dataset is obtained. This is the strategy adopted in   \textit{ab-initio} molecular dynamics\cite{Par-PRL-99} (AIMD) where the forces acting between molecules are computed at the QM level and the particles are subsequently moved by discretization of Newton's laws.  This is a powerful simulation technique but its numerical cost leads to several limitations. First, since it needs to solve QM equations multiple times, only the most efficient methods are appropriate. In practice it is essentially limited to the use of electronic density functional theory (DFT) to compute the forces. Second, the length and timescales accessible to AIMD are rather limited (typically up to a few $ns$ and  a few $nm$). Finally, high throughput calculations are not an option since each simulation requires $10^{4-5}$  CPU hours. 

When the solvent does not react with the solute, an usual strategy is to resort to classical force-field molecular dynamics (MD) to describe the solvent while keeping a QM description of the solute. This hybrid QM-MM\cite{LinTru-TCA-07} strategy  reduces tremendously the computational  cost. An even further simplification is to avoid the statistical sampling of the solvent degrees of freedom by modelling it  as a  continuum. This allows to account for part of solvation effect while maintaining the computational cost close to \textit{in vacuuo} calculations. It is thus a popular strategy  but it suffer from several limitations. Continuum solvation models (CSMs) have several parameters,  which are usually fitted to reproduce a set of reference calculations. Consequently the ab-inito nature of the approach is lost in the process, as well as the molecular nature of the solvent.

A third strategy in to conserve the QM-MM partitioning of the system but to resort to liquid state theories (LST) to describe the classical part. There exist several LST that aim at solving the Ornstein-Zernike (OZ) equation that is the fundamental equation describing liquids at the atomic/molecular scale. In the OZ equation, the liquid is described by a continuous field related to its spatially dependant density $n(\bm{r})$. Key quantities entering OZ equation are the direct and total correlation function $c$ and $h$ that account for the correlation within the liquid. These functions depend on the interaction potential between the liquid particles and the considered physical conditions, such as the pressure or temperature. It is worth insisting on the fact that such approaches truly conserves a molecular description of the fluid particles while relying on the rather simple continuous solvent density $n(r)$. This represents a compromise between the CSM, which fails to account for the molecular nature of the liquid, and the MD-based strategy, which relies on the tedious sampling of solvent configurations. The appealing 
properties of LST has motivated several attempts to utilize them for incorporating solvent effects in QM calculations. The RISM approach and its 3D-RISM counterpart are for instance implemented in several QM packages\cite{ten-no_hybrid_1993,naka_rism-scf_1999,casanova_evaluation_2007}. The idea behind 3D-RISM is to replace the molecular direct correlation function $c$  by a sum of simpler 1D site--site correlation functions. This is an nonphysical approximation that makes the theory inconsistent with statistical physics  principles but it nevertheless drastically simplifies the resolution of the molecular OZ equation. Another LST allowing to predict the in-homogeneous density of the liquid is  classical density functional theory (cDFT)\cite{mermin_thermal_1965,evans_nature_1979}. In short, cDFT allows to compute the grand-potential by minimizing a functional of the liquid density. This minimum is reached for the equilibrium density. 
Similarly to its electronic counterpart, cDFT is theoretically well grounded but  approximations in the choice of the functional  describing the interactions between particles, here solvent molecules, are required to make its utilisation practical. Several attempts to use cDFT as a solver for QM/MM calculations have been proposed. In the   joint DFT framework and its associated software JDFTx\cite{lischner_joint_2011,sundararaman_jdftx_2017},  model functionnals whose parameters are adjusted to account for properties of real fluid are proposed. In the so-called reaction density functional theory (RxDFT)\cite{tang_development_2018}, solute electronic structure is computed with  electronic DFT.  Atomic partial charges are fitted to best reproduce  the molecular electrostatic potential. These atomic charges, along with LJ (LJ) sites used to model dispersion and repulsion, are subsequently employed in a  classical DFT calculation to estimate the solvation free energy. 
A more elaborate strategy is to self-consistently minimize the electronic and the solvent functional to account for the mutual impact between the solute and the solvent. Even keeping the coupling Hamiltonian at the electrostatic level, it is possible to capture a large part of the mutual polarization between the solute and the solvent. This strategy has been employed to study the reactivity of an SN$_2$ nucleophilic substitution between chloride and chloromethane in water\cite{jeanmairet_tackling_2020}. In this original work, eDFT was chosen to describe the QM solute. 

Turning now to the theoretical description of spectroscopic properties in solution, 
the latter has focussed an increasing attention over the last thirty years. 
As for ground state or free energy calculations, there are mainly two approaches to describe the solvent: 
either the solvent is considered as a continuous medium 
\cite{KarZer-JACS-90,KarZer-92-JPC,MenCamTom-JCP-98,TomPer-CR-94,CosBar-JCP-00,BanGor-JCP-00,FatGyg-JCC-02,FatGyg-IJQC-03,ChiFedKit-JCC-08,IkeIshFedKitInaUmeYokSek-JCC-10,DziHelSkyMosPay-EL-11,HowWomDziSkyPriCra-JCTC-17},  
or explicitly treated as a molecular object either at the QM or MM level\cite{GorFreBanJenKaiSte-JPCA-01,KonOstKurAstChr-JCP-04,FusMinSliZahGor-JPCL-11}. 

Whether the solvent is treated at the continuum or molecular level, 
its effect on the spectrum of the Hamiltonian is obtained by adding an effective one-body potential  
accounting for the QM-MM interaction.  
Nevertheless, the main difference between the continuum and the explicit description from the point of view of the spectroscopy 
comes from the way the excited states properties are computed.  
In the case of the CSM, once the converged charge density of the dielectric is obtained, a single QM calculation is performed and correspondingly all the excited states properties. 
When the solvent is described using classical MD, once the solvent trajectory have been generated either using a quantum or a classical description of the solute, two strategies 
can be employed to compute the excited states. A first one is to determine the statistically averaged structure of the solvent and to use it as an the external field acting on the solute in a single QM calculation allowing to compute the excited states. Another one is to compute the excited states properties as a statistical average over a set of distinct QM calculations performed along the classical trajectory of the whole QM-MM system. The latter is of course more computationally demanding than the former since it requires to perform $10^2-10^3$ QM calculations.

Again, as in the  case of the ground state, there is a trade off between the accuracy of a molecular description and the appealing computational cost of a continuum description. 

In the present work, we propose to use the hybrid continuum/molecular description of the solvent 
offered by  molecular density functional theory (MDFT)  to compute excited state properties. 
Our approach consists in a self-consistent ground state MDFT calculation, which allows to obtain 
both the ground state and solvent equilibrium charge distribution. 
A  single excited states calculation in the presence of the electrostatic field 
of the equilibrium solvent distribution is subsequently performed. 
We tested the proposed methodology by studying the excited states properties of  water and formaldehyde in aqueous solution using near full-configuration interaction level of theory, namely converged selected configuration interaction (SCI)\cite{malrieu_cipsi,three_class_CIPSI,GinSceCaf-CJC-13,hbci,SchEva-JCTC-17,LooSceBloGarCafJac-JCTC-18,QP2-JCTC-19}, for the calculation of excited states,

We compare with state-of-the-art approaches using a coupled cluster (CC) level of theory for the QM part 
and a molecular\cite{KonOstKurAstChr-JCP-04} 
or continuum description\cite{Caricato-JCP-13,HowWomDziSkyPriCra-JCTC-17} of the solvent.  
For water in water, the results obtained with our QM-MMDFT approach are in better agreement with the reference values than the continuum approaches\cite{Caricato-JCP-13,HowWomDziSkyPriCra-JCTC-17}. A reasonable accuracy is also found for the formaldehyde in aqueous solution. 
We also investigate the impact of the level of theory used to describe the QM part in the MDFT calculation. To rationalize predicted excitation energies we also computed the  dipole moment of both ground and excited states, in vacuum and solution.
Contrary to previous claims \cite{Gonz-JPCB-18}, we found no clear evidence of a correlation between the intensity of the blue shifts and the dipole moments.

\section{Theory}
\indent

\subsection{MDFT as an effective point charges model for QM-MM calculations}
As the primary goal of this paper is to demonstrate that  MDFT is an appropriate method to account for solvent effects 
in QM calculations, we will expose the formalism from the QM perpective, \textit{i.e.}, as 
a self-consistent classical electrostatic embedding of the usual QM problem. 

In the present approach, the QM solute, treated within the Born-Oppenheimer approximation,  interacts with the solvent described as a volumetric  charge density $ \rho_c$. 
Therefore, the additional QM-MM potential $\vqmmm[\rho_c](\brr)$ is simply the Coulomb potential produced by $\rho_c$, which can be explicitly written as 
\begin{equation}
  \vqmmm[\rho_c](\brr) = \int\frac{\rho_c(\brr^\prime)}{|\brr'-\brr|} \text{d}\brr', 
\end{equation}
 using the standard atomic units, $1/4\pi \epsilon_0=1$. 
{Note that, as opposed to a continuous description, there is no relative permitivitty involved in the expression of the Coulomb potential but only the vacuum permittivity $\epsilon_0$}. 
In the present work, both the classical solvent charge density $\rho_c$ and the QM counterpart are obtained through the self-consistent resolution of the QM and MDFT equations, allowing to take into account the explicit coupling between the QM and MM part. 
Before describing in more details the MDFT equations in Sec. \ref{sec:MDFT}, we would like to emphasize that $\rho_c$ depends mainly on three quantities: the QM density, the choice of the classical functional describing the solvent and the description  of the QM-MM interaction \textit{i.e.} the coupling Hamiltonian. 

Although the classical density $\rho_c$ is in principle a continuous object, we represent it on a three-dimensional grid and therefore  {$ \vqmmm$} is obtained as a sum of point charges contribution
\begin{equation}
 \label{eq:vqmmm}
  \vqmmm[\rho_c](\brr) = \sum_{m=1}^{N_\text{g}} \frac{\rho_c(\brr_m)\Delta V}{|\brr_m-\brr|}, 
\end{equation}
where $\brr_m$ is the coordinate of the grid point $m$, $\Delta V$ the volume of the grid voxel, 
and $N_\text{g}$ the number of grid points. 

Given a classical charge density $\rho_c$, the total effective QM Hamiltonian $\hqmmm[\rho_c]$ is therefore 
the sum of the usual Born-Oppenheimer Hamiltonian and of the $\vqmmm[\rho_c]$ potential 
\begin{equation}
 \begin{aligned}
 \hqmmm[\rho_c] = \vnn + \hkin + \hvne + \hwee + \vqmmm[\rho_c]
 \end{aligned}
\end{equation}
where 
\begin{equation}
 \vnn = \sum_{A>B} \frac{\text{Z}_A \text{Z}_B}{|\brnucl{A} - \brnucl{B}|},  
\end{equation}
\begin{equation}
 \hkin = -\frac{1}{2} \sum_{i=1}^{N_e} \Delta_{\br{i}},  
\end{equation}
\begin{equation}
 \hvne = - \sum_{i=1}^{N_e} \sum_{A}\frac{Z_A}{|\br{i}-\brnucl{A}|},
\end{equation}
\begin{equation}
 \hwee = \sum_{i>j}\frac{1}{|\br{i}-\br{j}|}, 
\end{equation}
with $Z_A$ and $\brnucl{A}$ being the nuclear charge and coordinates, respectively, 
and $\br{i}$ the electronic coordinates. 
Developed on a AO basis set $\basis=\{ \chi_i(\brr) \}$, the classical Coulomb potential is then simply given as 
a sum of usual one-electron Coulomb integrals weighted by the classical charge density $\rho_c$ 
\begin{equation}
 \elemm{\chi_j}{\vqmmm[\rho_c]}{\chi_i}= 
  \sum_{m=1}^{N_g} \Delta V \rho_c(\brr_m)\elemm{\chi_j}{\frac{1}{|\brr_m-\brr|}}{\chi_i}. 
\end{equation}

For a given classical density of charge $\rho_c$, we can then solve the Schr\"odinger equation 
as usually done in any quantum chemistry calculations 
\begin{equation}
 \hqmmm[\rho_c] \ket{\Psi_i[\rho_c]} = E_i[\rho_c]\ket{\Psi_i[\rho_c]} \label{eq:Hdependon_rhoc}, 
\end{equation}
where the solutions $\ket{\Psi_i[\rho_c]}$ and $E_i[\rho_c]$ depend on the solvent charge density $\rho_c$. 
Therefore one can obtain any usual properties with  a computational cost similar to vacuum calculation 
while  explicitly taking into account the electrostatic effect of the solvent. 

In this paper, we shall be mostly concerned with electronic properties of both the ground 
and excited states, such as excitation energies 
\begin{equation}
 \omega_i[\rho_c] = E_i[\rho_c] - E_0[\rho_c], 
\end{equation}
and the dipole moments 
\begin{equation}
 {\boldsymbol \mu}_i[\rho_c] = {\boldsymbol \mu}_\text{nucl} - \int \text{d}\brr n_i(\brr) \brr,  
\end{equation}
where 
\begin{equation}
 {\boldsymbol \mu}_\text{nucl} = \sum_{A} Z_A \brnucl{A},  
\end{equation}
and $n_i(\brr)$ is the one-electron density of the state $\Psi_i[\rho_c]$, which depends implicitly on 
the classical charge density.  
The electronic wave-function depends explicitly on the solvent charge density as evidenced in Eq.~\eqref{eq:Hdependon_rhoc}. It will be shown in section \ref{sec:MDFT} that the solvent charge density is also influenced by the electronic density $n$.
The primary objective of this paper is to introduce a self-consistent QM-MDFT resolution to predict a reasonable solvent charge density and to compute the electronic structure properties in its presence.

\subsection{Summary of Molecular Density Functional Theory \label{sec:MDFT}}

Molecular density functional theory\cite{jeanmairet_molecular_2013,jeanmairet_tackling_2020,borgis_simple_2020} is a particular flavour of classical DFT that has been designed to tackle solvation problems. We shall start by recalling the main ansatz of classical DFT\cite{mermin_thermal_1965,evans_nature_1979}. We consider a system made of classical particles, \textit{i.e} whose interactions are described by a classical force-field, which is perturbed by the presence of an external potential $V_\text{ext}$. Following Mermin and Evans pioneering work, it is possible to show that in the grand-canonical ensemble, at finite temperature $T$, there exists a unique functional $\Omega[\rho]$ of the particle density, $\rho$. This functional is equal to the grand potential $\Omega$ at its minimum, which is reached for the equilibrium solvent density $\rho_{\text{eq}}$. In the framework of MDFT, the external potential is due to the presence of the solute and the particle density describes the solvent. We will consider solvent constituted of rigid molecules such as the knowledge of the position of their center of mass $\bm{r}$ and  orientation 
$\bm{\Omega}$ is sufficient to fully describe molecule coordinates. Consequently, the solvent density $\rho(\bm{r},\bm{\Omega})$ have a spatial and angular dependancy. 
We introduce the functional $F$, defined as the difference between the grand potential functional in the presence of solute $\Omega[\rho]$ and the grand potential $\Omega_0$ of the non perturbed fluid, \textit{i.e} with the homogeneous density $\rho_0$,
\begin{equation}
  F[\rho]=\Omega[\rho]-\Omega_0.
  \label{F=Omega-Omega0}
\end{equation}
The minimization of this functional directly yields the solvation free energy (SFE) and the equilibrium solvent density. It is noteworthy to emphasize the challenge associated with computing SFE using methods based on statistical sampling of the phase space, such as MD. This difficulty arises because the free energy cannot be expressed as an ensemble average but instead necessitates knowledge of the grand-partition function\cite{frenkel_understanding_2002}.

The functional $F$ is usually written as 
\begin{equation}
  F[\rho]=F_{\text{id}}[\rho]+F_{\text{exc}}[\rho]+F_{\text{ext}}[\rho].
  \label{F=Fid+Fexc+Fexc}
\end{equation}
In this equation, the first term $F_{\text{id}}$ represents the ideal term, corresponding to the entropic contribution of the non-interacting fluid\cite{evans_nature_1979,hansen_theory_2006}. The second term, $F_{\text{exc}}$, is due to solvent-solvent interactions. Similarly to the exchange-correlation term in electronic DFT, this is the problematic term. Note that, unlike electronic DFT, the excess functional is not unique, as it depends on the force-field model chosen to describe the interaction between particles. While an analytical expression exists, it requires approximations to be used in practice. In this work, we use the so-called HNC excess functional $F_{\text{exc}}$\cite{ding_efficient_2017},
\begin{equation}
    F_\text{exc}[\rho]=-\frac{k_BT}{2}\iint \Delta \rho(\bm{r}) c(|\bm{r}-\bm{r^}\prime|,\bm{\Omega},\bm{\Omega}^\prime)  \Delta \rho(\bm{r}^\prime) d\bm{r}d\bm{r}^\prime d\bm{\Omega}d\bm{\Omega}^{\prime} \label{eq:Exc}
\end{equation}
where $k_\text{B}$ is the Boltzmann constant and $\Delta \rho=\rho-\rho_0$.
The remaining term, $F_\text{ext }$, represents contribution due to the external perturbation $V_\text{ext}$ of the solute acting on the solvent. It can be formally expressed as
\begin{align}
 F_\text{ext} [\rho] &= \iint\rho(\bm{r}, \bm{\Omega}) V_\text{ext}(\bm{r},\bm{\Omega}) d\bm{r}d\bm{\Omega}    \label{F_ext}  \\
                      &=  F_\text{ext,C} [n, \rho] + F_\text{ext,RD} [\rho,n] \nonumber
\end{align}
which can be split into the electrostatic contribution $F_\text{ext,C} $ and the repulsion-dispersion contribution $F_\text{ext,RD}$. The electrostatic interaction between the quantum solute and the classical solvent , which depends on the electronic density $n$ is
\begin{equation}
  F_\text{ext,C} [n, \rho] = \int \rho_c(\bm{r}) V_\text{QM}[n](\bm{r}) d\bm{r}.
  \label{eq:F_extq}
\end{equation}
where $V_\text{QM}[n]$ is  the electrostatic potential generated by the nuclei and the solute electronic density. The  solvent charge density can be computed from the solvent density with
\begin{equation}
  \rho_c(\bm{r})= \iint \rho(\bm{r}', \bm{\Omega}) \gamma(\bm{r} - \bm{r}',\bm{\Omega}) d\textbf{r}'d\bm{\Omega},
  \label{charge_dens_solv}
\end{equation}
where $\gamma(\bm{r},\bm{\Omega})$ is the charge density of a water molecule taken at the origin with orientation $\bm{\Omega}$. 
In addition to the electrostatic term of Eq.~\eqref{eq:F_extq}, one should also include short-range repulsion and dispersion interactions between the solute and the solvent. Using a popular strategy in QM/MM literature\cite{freindorf_marek_optimization_1998}, this is done by adding  LJ (LJ) sites on the nuclei of the solute. The repulsion-dispersion part of the functional then takes the following form
\begin{equation}
  F_\text{ext,RD} [\rho,n] \approx F_\text{ext,LJ} [\rho] = \iint \rho(\bm{r}, \bm{\omega}) V_\text{LJ}(\bm{r},\bm{\omega}) d\bm{r}d\bm{\Omega},
  \label{F_extLJ}
\end{equation}
with
\begin{equation}
\begin{split}
\quad V_\text{LJ} (\bm{r}, \bm{\Omega}) = &\sum_{i \in \text{solute} } \sum_{j \in \text{solvent} } 4 \epsilon_{ij} \left[ \left(\dfrac{\sigma_{ij}}{\abs{\bm{r} + \bm{r}_{j\bm{\Omega}} - \bm{r}_i}}\right)^{12}\right.\\
&
-\left. \left(\dfrac{\sigma_{ij}}{\abs{\bm{r} + \bm{r}_{j\bm{\Omega}} - \bm{r}_i}}\right)^6 \right].
\end{split}
\label{V_LJ}
\end{equation} 
where  $\epsilon_{ij}$ and $\sigma_{ij}$ are the mixed LJ parameters. In Eq.~\eqref{V_LJ}, $\bm{r}_i$ is the position of the i$^\text{th}$ site of the solute and
$\bm{r}_{j\bm{\Omega}}$ denotes the position with respect to the center of mass of the $j^\text{th}$ site of a solvent molecule located in $\bm{r}$ and having an orientation $\bm{\Omega}$.
There is no prescribed method to choose the LJ parameters but it impacts the predicted solvation free energies and structure \cite{jeanmairet_tackling_2020}. As a result, the proposed methodology is no longer truly ab initio. An alternative could be to include repulsion and dispersion interactions through the use of an electron-solvent pseudopotential\cite{turi_analytical_2001,marefat_khah_avoiding_2020}. 
The QM and MM part are thus only coupled at the electrostatic level through Eqs.~\eqref{eq:vqmmm} and \eqref{eq:F_extq}, which should in principle give the same result if no approximations were made in their numerical implementation. 
In the next subsection, we detail how the joint resolution of Eqs.~\eqref{eq:Hdependon_rhoc} and \eqref{F=Fid+Fexc+Fexc} are performed to self-consistently compute the electronic density and the solvent charge density.

\subsection*{C. QM/MM Coupling}
\indent

The first iteration of our framework begins at the quantum level by computing the energy of the solute in vacuum.  The electrostatic potential $V_\text{QM}[n]$, which depends on the ground-state electronic density is subsequently employed in the calculation of the external functional of Eq.~\eqref{F_ext} in the MDFT calculation. After optimization of the classical functional, the  solvent charge density computed as in Eq.~\eqref{charge_dens_solv} is introduced as a set of point charges in a new QM calculation as in Eq.~\eqref{eq:vqmmm}. This allow to obtain the ground-state energy of the solute in solution. This process is iterated until convergence is achieved, that is  when the solute energy variation is below $5\cdot10^{-4}$ a.u and the classical free energy variation is below $10^{-4}~\text{kJ}\cdot \text{mol}^{-1}$ in this paper. 


The introduced QM-MDFT properly accounts for a large part of the mutual polarization of the electronic density and the molecular density of the solvent. Moreover, solvation structure is predicted in great details since the equilibrium density $\rho({\boldsymbol{r},\boldsymbol{\Omega}})$ measures the average number of molecules at specific locations but also quantifies the preferred  orientations. 
Nevertheless, as this paper exclusively focuses on a non-polarizable water model, it overlooks the electronic polarizability of the solvent molecules. This is a strong limitation since the  electronic density of the solvent relaxes almost instantaneously upon a modification of the electric field. 

\section*{3. Computational details}
\indent

This article focuses on the aqueous solvation of two polar solutes: water and  formaldehyde. 
Except for equation of motion coupled cluster restricted to single and double substitution (EOM-CCSD )calculations which have been performed with PySCF\cite{sun_recent_2020},
all QM calculations are performed using the  Quantum Package\cite{QP2-JCTC-19}.  
The electrostatic potential created by the solute is evaluated in a  25$^3$ \AA$^3$ cubic box with 100 grid nodes in each direction, and serves as an external potential in the following MDFT calculation.

Solvent densities are obtained using our homemade Fortran-written MDFT program. Minimizations are performed using the same spatial grid as the one used to compute the QM electrostatic potential, and the three Euler angles are described with 196 discrete orientations.
 Water solvent molecules are described using the non-polarizable  SPC/E model \cite{berendsen_missing_1987}. The $c^{(2)}$ direct correlation function, entering the excess functional of Eq.~\eqref{eq:Exc}  has been kindly provided by Luc Belloni. He computed it for an homogeneous density $n_0=0.0332891\ $\AA$^{-3}$ using a mixture of Monte-Carlo simulation data and integral equation theory \cite{belloni_exact_2017}.  As described earlier in the text,  dispersion and repulsion forces between the solute and the solvent molecules are modelled by adding LJ sites onto the solute nuclei. We use the same LJ parameters as the QM/MD we compare to \cite{kongsted_dipole_2002,kongsted_linear_2003,KonOstMikChr-JPCA-03,osted_statistical_2006,HowWomDziSkyPriCra-JCTC-17}. When comparing to CSM results which do no need LJ sites, we  take our parameters from the second generation generalized amber force field (gaff2)\cite{he_fast_2020}.

The proposed QM/MDFT procedure can be used to predict the to equilibrium solvent density from which the solvent charge 
density can be deduced from Eq.~\eqref{charge_dens_solv}. As mentioned earlier, the choice of the QM method 
used to obtain the ground state electronic density impacts the predicted solvent charge density. 
To assess the impact of this choice, three QM methods were considered: Hartree-Fock (HF), 
Configuration Interaction Single Double (CISD) and SCI\cite{QP2-JCTC-19}. 
Nevertheless, the excitation energies and dipole moment  are always computed using SCI, regardless of the QM/MDFT method employed, in order to avoid any approximations on these subtle properties.

\section{Applications}

\subsection{Water in Water}
\label{sec:wat_1}
We start by the investigation of the three lowest singlet excited states of  water, both in vacuum and in solution. 
This system has been studied previously using various techniques. 
In particular, two research groups used the equation of motion coupled cluster 
restricted to single and double substitution (EOM-CCSD)  to describe the solute
\cite{kongsted_dipole_2002,kongsted_linear_2003,KonOstMikChr-JPCA-03,osted_statistical_2006,HowWomDziSkyPriCra-JCTC-17}.
However, these studies relied on two different approaches to account for the solvent effects. The group of Christiansen  used a MD approach in which they first run a fully classical simulation to generate a set of relevant solvent configurations that are subsequently use as an external potential in an EOM-CCSD calculation. In practice, they assigned partial point charges to the
MM nuclei which are introduced into the one-electron part of the vacuum Hamiltonian. They still resort to LJ sites on the quantum nuclei to account for Pauli repulsion and dispersion. 
Then, they  either perform a single QM calculation on a configuration representative of the average over the trajectory\cite{kongsted_dipole_2002,kongsted_linear_2003,KonOstMikChr-JPCA-03}. Another strategy is to perform a series of several hundreds of QM calculation with a set of solvent configuration selected along the MM trajectory\cite{osted_statistical_2006}. The first strategy is an  equilibrium approach and closely resembles the MDFT-based methodology proposed here. The average solvent configuration employed is nothing more than some sort of projection of the equilibrium density onto the most representative  set of coordinates. 
In contrast, the second approach encompasses some dynamical information as it properly accounts for the fluctuations of the solvent. This, however, comes at the cost of a drastic increase in computational needs.

On the other hand, Howard \textit{et. al.} \cite{HowWomDziSkyPriCra-JCTC-17}  used an implicit solvent model based on a inhomogeneous permittivity\cite{fattebert_density_2002,scherlis_unified_2006,andreussi_revised_2012} which depends locally on the electronic density. The dielectric constant transition smoothly between $\epsilon=1$ for high density region, i.e. close to solute, towards the permittivity of the homogeneous fluid far from it. 

\subsubsection{Benchmark of water in vacuum}
\label{sec:water_vac}
Our primary goal is to be able to compare our QM/MDFT results with  existing data.  However, different studies often use different  geometries and  basis sets. Therefore, in order to allow for a fair comparison, we start  by
discussing how  the geometry, the basis set  and the level 
of theory  influence the predicted excitation energies in vacuum.

We start by assessing the impact of the solute coordinates by comparing the predictions using the geometries reported by Kongsted  \textit{et. al.}\cite{kongsted_dipole_2002} and Howard \textit{et. al.}\cite{HowWomDziSkyPriCra-JCTC-17}.
In Table \ref{Water_exc_nrg_vac_tot} we report the excitation energies computed both at the EOM-CCSD and SCI level of theory, in the Dunning aug-cc-pVXZ basis sets.
In Kongsted's geometry, the OH bond length is about 0.01 \AA \ larger and the HOH angle is wider by roughly $1^\circ$ with respect to Howard's.
From Table \ref{Water_exc_nrg_vac_tot} we can notice observe that the geometry plays a substantial role because, despite their similarities, 
the excitation energies computed using Howard's geometry are between 0.04 eV to 0.08 eV lower than the values obtained using Kongsted's. 
This is a significant differencesince our target accuracy  is about 0.04 eV (\textit{i.e.} 1 kcal/mol).

\begin{table}[t]
\centering                           
\begin{tabular}{c||cc|cc|cc||cc|cc}
                              &  \multicolumn{6}{c||}{\textbf{EOM-CCSD}}                       &  \multicolumn{4}{c}{\textbf{SCI}} \\ \hline \hline
\multicolumn{1}{c||}{\textbf{State}}   &\multicolumn{2}{c|}{\textbf{AVDZ}}&\multicolumn{2}{c|}{\textbf{AVTZ}} & \multicolumn{2}{c||}{\textbf{d-AVTZ}}  &\multicolumn{2}{c|}{\textbf{AVDZ}} &\multicolumn{2}{c}{\textbf{AVTZ}}                           \\ \hline
\multicolumn{1}{c||}{$\bm{^1B_1}$ }&        7.38 & 7.46      & 7.54 & 7.61              & 7.53  & 7.62                 &   7.47  & 7.54           & 7.56  & 7.64                                        \\ \hline
\multicolumn{1}{c||}{$\bm{^1A_2}$} &        9.16 & 9.22      & 9.31 & 9.37              & 9.30  & 9.37                 &   9.26  & 9.33           & 9.35  & 9.42                                        \\ \hline
\multicolumn{1}{c||}{$\bm{^1A_1}$ }&        9.82 & 9.87      & 9.93 & 9.96              & 9.84  & 9.88                 &   9.92  & 9.94           & 9.96  & 9.99               
\end{tabular}
{\caption{Excitation energies (eV) for water in vacuum computed at the EOM-CCSD and SCI level of theory using the geometry of Howard \textit{et. al.} (left) and the geometry of Kongsted \textit{et. al.} \cite{KonOstMikChr-JPCA-03} using various  Dunning basis sets.}\label{Water_exc_nrg_vac_tot}} 
\end{table}

Interestingly, the changes in excitation energy are very similar for both geometries when changing the basis set. 
We will thus focus on Howard's geometry  to discuss the influence of the basis set, as similar conclusions can be drawn with Kongsted's geometry.
Within the EOM-CCSD framework, the changes in excitation energy between the aug-cc-pVDZ and aug-cc-pVTZ basis sets are 0.16 eV, 0.15 eV and 0.11 eV for the first, second and third excited state, respectively, 
while using the SCI level of theory they are slightly smaller (\textit{i.e.} 0.09 eV, 0.09 eV, 0.04 eV, respectively).  
Nevertheless, using the EOM-CCSD method, the change in excitation energies between the aug-cc-pVTZ and d-aug-cc-pVTZ basis sets is much smaller 
(\textit{i.e.} 0.01 eV, 0.01 eV and 0.09 eV, for the first, second and third excited states, respectively), and we can expect similar trends 
at the SCI level of theory.  
We can conclude that an aug-cc-pVTZ basis set is sufficient for the description of the first and second excited state, while the aug-cc-pVDZ basis set is sufficient for the third excited state as error cancellations lead to an excitation energy close to the d-aug-cc-pVTZ description. 

Eventually, we can compare the results obtained with different levels of theory, \textit{i.e.} EOM-CCSD and SCI, keeping in mind that our converged 
SCI results converge towards the Full Configutation Interaction (FCI) and therefore are the reference within a given basis set. 
From the results of Table \ref{Water_exc_nrg_vac_tot} we can notice that, in the aug-cc-pVDZ basis set, 
the EOM-CCSD method systematically underestimates the excitation energies by typically the same amount for both geometries, 
which ranges between 0.07 eV to 0.11 eV. Interestingly, when using the aug-cc-pVTZ basis set, the bias of EOM-CCSD excitation energies 
with respect to SCI is significantly lowered, ranging from 0.02 eV to 0.05 eV.

To conclude this part, it appears clearly that the geometry, the QM method and the basis set can all together have a significant 
influence on the predicted excitation energy. 
For instance, the first excitation energy differs by about 0.3 eV between the EOM-CCSD in the aug-cc-pVDZ 
basis set using Howard's geometry and the SCI in the aug-cc-pVTZ basis set using Kongsted's geometry. 
Regarding both the choice of the QM method and the basis set, it is of course driven by computational resources considerations. 
While in the case of a small system such as the water molecule we could obtain near FCI calculations, it becomes soon impossible when increasing the system size, even with efficient SCI algorithms. 
In the case of the water molecule, we can safely estimate that the EOM-CCSD in the aug-cc-pVTZ is within 0.05 eV to FCI, 
and that the use of larger basis set only impacts the third excited state by about 0.1 eV. 
This allows to estimate the first source of error between our framework and the literature. 
Finally, the choice of the geometry may have the least straightforward effect to rationalize,  
as even a slight change in geometry can impact by about 0.08 eV the predicted excitation energies. 
This can become problematic since while it is rather straightforward to define a geometry in vacuum 
- for instance by geometry optimization -, this not the case  in solution. Indeed, the geometry of the solute is most likely fluctuating due tho thermal agitation. It is crucial to keep this in mind when comparing theoretical predictions to experimental data. 

\subsubsection{Water in water}

We now turn to the study of water, described at the QM level, solvated into liquid water, described at the MDFT level using the methodology described above. An important point to remember is that the solvent is not allowed to relax during the excitation process. Consequently, even in the computation of the excited states, the solvent charge density is the one in equilibrium with the ground-state electronic density. 
Thus, regardless of the choice of the QM method used to compute the excitation energies (e.g., SCI in our case), the solvatochromism ultimately depends on the classical solvent charge density obtained through the self-consistent QM-MDFT procedure.

Since the QM-MDFT procedure involves multiple QM and MDFT serial calculations, 
it is  worth investigating how the predicted excitation energies are impacted by the level of theory used to describe the ground state electronic density in this process.

To address this question, three wave function methods of increasing level of complexity 
were employed to predict the solvent structure with the QM-MDFT procedure: HF, CISD and SCI. 
Each QM method produces a different solvent density, which are labeled $\rho_0$, $\rho_2$ and $\rho_\infty$, respectively,  
according to the highest level of excitation considered in the method.

The excitation energies computed with SCI in the aug-cc-pVDZ basis set using the equilibrium solvent densities are reported in Table \ref{Water_exc_nrg_sol} . 

\begin{table*}
\center
\begin{tabular}{c||c|c|c||c}
          & \multicolumn{3}{c||}{\textbf{QM/MDFT}} & \textbf{Vacuum}  \\ \hline \hline
{\textbf{State}}   & $\bm{\rho_\infty}$ & $\bm{\rho_2}$   & $\bm{\rho_0}$ &        \\
\hline
{\bm{$^1B_1$}} &  8.04/8.12    &  8.05/8.12   & 8.08/8.15  & 7.47/7.54       \\    \hline                                                           
{\bm{$^1A_2$}} &  9.94/10.00    &  9.95/10.00  & 9.98/10.03  & 9.26/9.33      \\ \hline
{\bm{$^1A_1$}} &  10.42/10.46   &  10.43/10.46 & 10.46/10.49 & 9.92/9.94       \\
\end{tabular}
\caption{H$_2$O, aug-cc-pVDZ: excitation energies computed with SCI in vacuum and in solution using the solvent charges obtained with our QM-MDFT approach. 
The solvent density obtained using HF, CISD and SCI to described the ground state electronic density in the QM-MDFT procedure are denoted as $\rho_0$, $\rho_2$ and $\rho_\infty$ respectively. 
Calculations are reported with the geometry of Howard \textit{et. al.} (left) and that of Kongsted \textit{et. al.} (right). }
\label{Water_exc_nrg_sol}
\end{table*}

As appears in Table \ref{Water_exc_nrg_sol}, accounting for the solvent environment results in a notable increase in the excitation energies. This is true for both the geometries and for all three QM level of description tested to describe the ground state density during the QM-MDFT calculation.
This observation is relatively easy to rationalize: since the solvent density is optimized for the ground-state electronic density, it tends to stabilize the ground-state energy more than the excited states, leading to a blue-shift in the excitation energies.

More importantly, the excitation energies exhibit a weak dependency on the chosen solvent density, with the observed deviation being at most 0.04 eV (less than 1 kcal/mol).
The latter is very encouraging because it means that, at least for this system, performing the QM-MDFT optimization at the simple 
HF level is sufficient to determine the solvent equilibrium density, which can subsequently serve in more elaborate quantum methods. 
We can also remark that the excitation energies computed using $\rho_\infty$ are lower than the ones computed using $\rho_2$, which are themselves lower than the ones computed using $\rho_0$.
This can be rationalized by noticing that the permanent dipole of the water molecule, computed in vacuum and reported in Table \ref{dip_vac}, follows a similar trend: $\mu_{\text{HF}}>\mu_{\text{CISD}}>\mu_{\text{SCI}}$. 

\begin{table}[h!]
\center
\begin{tabular}{c||cc|c}
\multicolumn{1}{c||}{\textbf{QM Method}}     & \multicolumn{1}{c|}{\textbf{SCI}}   & \multicolumn{1}{c|}{\textbf{CISD}}  & \textbf{HF} \\ \hline \hline
\multicolumn{1}{c||}{$\bm{\mu}$} & \multicolumn{1}{c|}{0.732} & \multicolumn{1}{c|}{0.745} & 0.796 
\end{tabular}
\caption{Dipole moments $\mu_i$ (u.a)  of the ground state of the water molecule, computed  in vacuum using HF, CISD \& SCI with the aug-cc-pVDZ basis set. Calculations were made for Howard's geometry.}
\label{dip_vac}
\end{table}


Therefore, a more polar solute tends to induce larger polarization in the surrounding solvent, which itself leads to stronger stabilization of the ground state compared to the excited states. This justifies the observed trend where both the dipole moment and the excitation energies follow the same pattern.

We also report in Table \ref{dip_h2o} the ground state and excited states dipole moments of the water molecule in solution, computed at the SCI level. 
As observed in Table \ref{dip_h2o}, the dipole moment of the ground state increases in solution with respect to vacuum, as expected due the polarisation of the QM solute by the polar solvent. The strength of the ground state dipole moment of the water molecule follows a  trend similar to the one observed for the excitation energies: $\mu_\text{GS}[\rho_0]>\mu_\text{GS}[\rho_2]>\mu_\text{GS}[\rho_\infty]$. This is again a consequence of  the fact that the solvent density predicted with HF, $\rho_0$,  is more polarized than the densities predicted with the correlated methods $\rho_2$ and $\rho_\infty$.

\begin{table}[h!]
\center
\begin{tabular}{c||c||ccc}
 & \textbf{Vacuum}  & \multicolumn{3}{c}{\textbf{Solvent}}\\ \hline \hline

\multicolumn{1}{c||}{\textbf{State}} &  & \multicolumn{1}{c|}{$\bm{\rho_\infty}$} & \multicolumn{1}{c|}{$\bm{\rho_2}$} & $\bm{\rho_0}$ \\ \hline
\multicolumn{1}{c||}{\textbf{GS}} & 0.732 & \multicolumn{1}{c|}{0.880} & \multicolumn{1}{c|}{0.882} & 0.891 \\ \hline
\multicolumn{1}{c||}{$\bm{^1B_1}$} & -0.548 & \multicolumn{1}{c|}{-0.028} & \multicolumn{1}{c|}{-0.020} & 0.021 \\ \hline
\multicolumn{1}{c||}{$\bm{^1A_2}$} & -0.397 & \multicolumn{1}{c|}{-0.083} & \multicolumn{1}{c|}{-0.078} & -0.054 \\ \hline

\multicolumn{1}{c||}{$\bm{^1A_1}$} & -0.434 & \multicolumn{1}{c|}{0.074} & \multicolumn{1}{c|}{0.082} & 0.120 

\end{tabular}
\caption{Dipole moments  (u.a) of the ground sate and three first excited state of the water molecule in vacuum and water. All calculations were performed at the SCI level. Calculations in solution were performed with the solvent charge density generated  using  different QM methods: HF, CISD \& SCI.Calculations were made for Howard's geometry using the aug-cc-pVDZ basis set.}
\label{dip_h2o}
\end{table}

Regarding now the dipole moments of the excited states, one can first notice that in vacuum they have an opposite sign with respect to the ground state. The latter indicates a flipping of the charge distribution during excitation. 
In the presence of the solvent, one can notice that, with respect to the vacuum calculations, the dipole moments of the excited states are much lower in absolute value: they tend to align with the ground state dipole. 
This can be explained since the  solvent density is optimised for the ground state charge distribution, it therefore favors dipole moments aligned with the one of the ground state. The excited states have then to adapt to this surrounding electrostatic potential and tend to flip. 

\begin{figure}[b]
    \centering
    \includegraphics[width=0.95\linewidth]{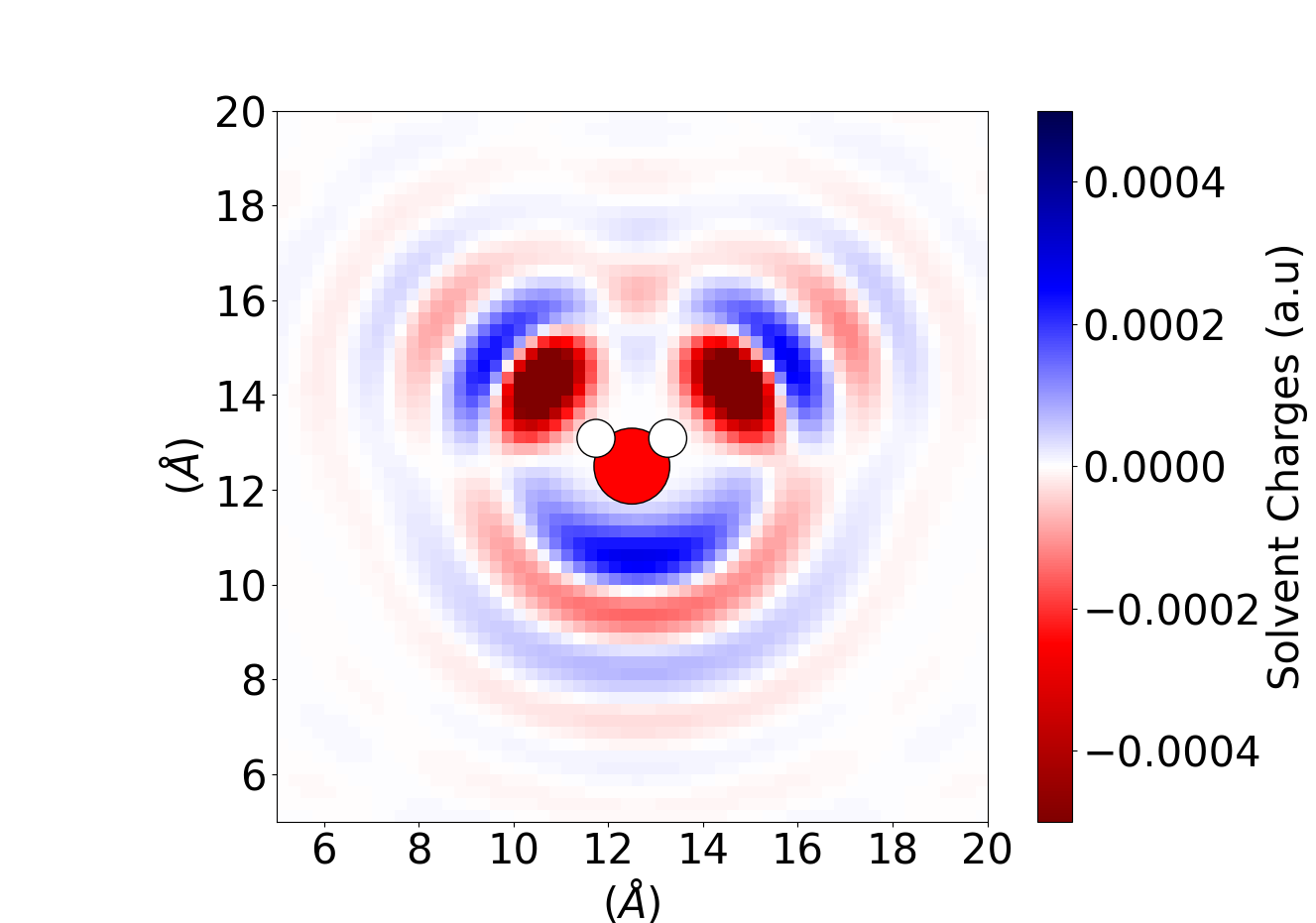}\\
    \includegraphics[width=0.95\linewidth]{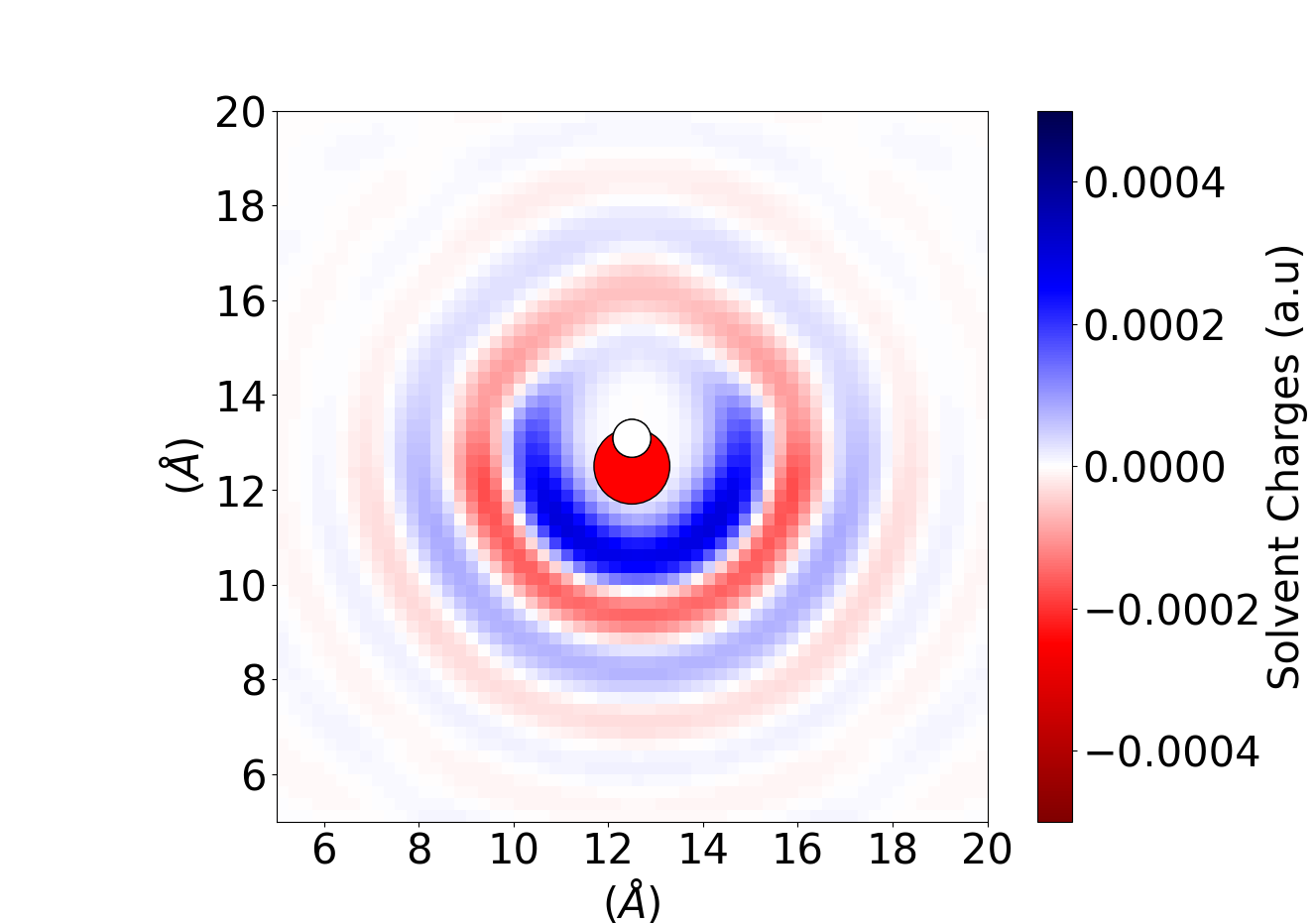}
    \caption{Solvent charges (u.a) in the plane of the molecule (upper) and in the bisector plane (lower). Calculations were made for Howard's geometry using the aug-cc-pVDZ basis set. CISD has been used to describe the electronic density during the QM-MDFT procedure. }
    \label{fig_h2o_dens}
\end{figure}

The tendency of the solvent to favor dipole moments aligned along the O-H direction is illustrated in a pictorial way in Figure \ref{fig_h2o_dens} where the charge density of the solvent surrounding the solute is displayed in the parallel and the bisector plane of the molecule. The first solvation shell is characterised by a positive charge surrounding the oxygen atom while a negative charge surrounds the hydrogen centers.  The charged regions in the immediate vicinity of the solvent are surrounded by regions of opposite charges, which is typical of a polar solvent. The second and third solvation shells are also captured, as evidenced by the oscillating behavior of the charge distribution. 
Overall, MDFT is able to predict the non-trivial three-dimensional structure of the solvent, as shown by the two slices displayed in Figure \ref{fig_h2o_dens}. In particular, the tetrahedral shape of the charge distribution, resulting from hydrogen bonding, is well rendered. This is a clear advantage of  MDFT as compared to simpler CSM, which do not capture the molecular nature of the solvent but instead describe it as a continuum.

We now compare the excitation energies predicted with the proposed SCI-MDFT framework to the results obtained  by Howard \textit{et al.}   using  PCM-EOM-CCSD and by Kongsted \textit{et al.} using  the CCSD-MM. 
We report in Table \ref{Water_exc_nrg_sol_all} the excitation energies obtained with SCI-MDFT in the aug-cc-pVDZ and aug-cc-pVTZ basis sets for  water in an environment described by the solvent density $\rho_2$, \textit{i.e} a density generated by a solute description at the CISD level.
We performed computations with both the geometry of Howard \textit{et. al.} and Kongsted \textit{et. al.}

A first observation is that our predicted excitation energies are consistently higher than the one obtained using the CSM, by a 0.2 to 0.4 eV. However, the agreement with the explicit modelling of the solvent is much better. This could be related to the similar level of description in MD and MDFT. The water model employed in MDFT being non-polarizable, we start by comparing to the non-polar CCSD-MM description. We can notice that we underestimate the first and third excitation energy by 0.02 eV and 0.15 eV, respectively while overestimating 
the second excited state by 0.08 eV. 
Overall, the agreement is quite satisfactory considering the numerous differences between the two calculations. For instance, the QM methods differ, which might induce \textit{per se} a difference of about 0.05 eV, as mentioned in Sec.~\ref{sec:water_vac}.
Also, the classical water force field are different, and, finally the excess functional of Eq.\eqref{eq:Exc} is approximated and therefore 
the solvent density is not exact. 

Interestingly, the overall agreement with Kongsted's predictions including polarization effect is  better, which can likely be attributed to error cancellations.

To conclude this part, we have demonstrated that the proposed QM-MDFT methodology enables the computation of excitation energies and dipole moments. These computations are in good agreement with state-of-the-art computations, in the case  of the water molecule.
Moreover, the main trends of "blue" solvatochromism can be relatively well rationalized with simple considerations based 
on electrostatic and dipole moments. 
From a computational perspective, an important finding is that performing the self-consistent QM-MDFT optimization with a low-level QM method, such as CISD or even HF, does not significantly alter the predicted excitation energies.

\begin{table}[h!]
\center
\begin{adjustbox}{width=0.5\textwidth}
\begin{tabular}{c||c|c||c|c|c||c}
          & \multicolumn{2}{c||}{\textbf{AVDZ}}    & \multicolumn{3}{c||}{\textbf{AVTZ}} & \multicolumn{1}{c}{\textbf{experiment}\cite{kerr_optical_1972}}   \\    \hline \hline                                    {\textbf{States}}         & \multicolumn{1}{c|}{$\bm{\rho_2}$$^a$}  & {\textbf{PCM}}     & \multicolumn{1}{c| }{$\bm{\rho_2}$$^a$} & {\textbf{PCM}} & \textbf{MD}(pol/nopol)$^b$   & \multicolumn{1}{c} {} \\ \hline
{$\bm{^1B_1}$} & {8.05}/{8.12}   & {7.82}  & 8.15/{8.22}  & {7.99} & 8.18/8.24  & 8.2  \\ \hline
{$\bm{^1A_2}$} & {9.94}/{10.00}  & {9.56}  & {10.03}/10.10 & {9.72} & 9.97/10.02 & XX   \\ \hline
{$\bm{^1A_1}$} & {10.43}/{10.46} & {10.16} & {10.48}/10.51 & {10.27}& 10.56/10.66& 9.9   
\end{tabular}
\end{adjustbox}
\caption{Excitation energies $\Delta E_i$ (u.a) for the water molecule in spce water solvent in aug-cc-pVDZ and  aug-cc-pVTZ  basis sets compared with results from the literature. \\
$^a$: results (left) are obtained with the geometry of Howard while results (right) are obtained with the geometry of Kongsted. \\
$^b$: results from Kongsted \textit{et. al.} using an d-AVTZ basis set with a polarizable or non polarizable force-field.  }
\label{Water_exc_nrg_sol_all}
\end{table}

\subsection{Formaldehyde in Water}
We now turn to the study of excitation properties of formaldehyde, a small organic molecule, in aqueous solution. This choice was partly guided by the fact that it was also studied by  Kongsted  \textit{et. al.}\cite{KonOstKurAstChr-JCP-04} and Howard  \textit{et. al.}\cite{HowWomDziSkyPriCra-JCTC-17} using the same methodology as the one applied to the study of water. Here again, Kongsted and Howard's geometry slightly differs, but in both cases the molecules are planar with a C$_2$ symmetry axis along the CO bond. Notably, Kongsted \textit{et. al.} only studied the first excited state of formaldehyde, which corresponds to the $n\rightarrow \pi^*$ transition while Howard  \textit{et. al.} computed the 8 first excited states. Our study focused on the four first excited states. Similarly to the study of water, several factors may impact the predicted excited states, such as the choice of the geometry and the LJ parameters used to model dispersion and repulsion, the choice of the solvent model and the level of theory employed to describe the quantum solute. For the sake of compactness, we will limit ourselves to the geometry of Howard  \textit{et. al.}.
Regarding the choice basis set, Loos \textit{et. al.}\cite{LooSceBloGarCafJac-JCTC-18} have shown that going from aug-cc-pVDZ to aug-cc-pVTZ change the excitation energies of the four first excited states of formaldehyde by at most 0.12 eV when using SCI. Since these differences in energy are comparable to the effect of geometry modification, there is no interest to go beyond aug-cc-pVDZ in our calculations. 
The excitation energies computed in vacuum are reported in Table \ref{tab:form_exc_vac}, where they are compared to those obtained by Howard  \textit{et. al.} and Kongsted \textit{et. al.}. 
Comparing the EOM-CCSD results of Howard et al. with our SCI results, we observe a maximum deviation of 0.04 eV for the third excited state, 0.01 eV for the fourth excited state, and below 0.01 eV for the first two excited states. These observations indicate that the EOM-CCSD approach does not introduce bias from the electronic structure for these states. Additionally, when comparing the EOM-CCSD results of Howard \textit{et al.} with those of Kongsted \textit{et al.}, we observe that the impact of changing the geometry on the first excitation energy is only 0.03 eV, suggesting a weak effect.

We now focus on the spectroscopic properties of formaldehyde in aqueous solution. The excitation energies computed using the SCI-MDFT procedure with the solvent charge density predicted using HF, CISD and SCI for the aug-cc-pVDZ basis set are compared to that of Howard \textit{et. al.} and Kongsted \textit{et. al.} in Table \ref{tab:form_exc_solv}. 
The choice of the QM method used to predict the solvent charge has little impact on the excitation energies of the two first excited states, as the energies differences are no greater than 0.04 eV. 
The impact on the higher excited states is more marked, reaching up to 0.18 eV for the fourth excited state. This might be explained by the  fact third and fourth excited states correspond to very diffuse Rydberg states.  
For the lowest excited state, our computed excitation energies are systematically underestimated with respect to EOM-CCSD/CSM and EOM-CCSD/MD which are consistently larger by roughly 0.2 eV. Again, several approximations can be the cause of this underestimation but we believe that the lack of polarization of our solvent model could play the major role.
For the three following excited states, our SCI-MDFT predicts a significant increase in the excitation energies with respect to vacuum. In contrast, Howard \textit{et. al.} surprisingly report almost no  solvent effect using EOM-CCSD/CSM. 
Unfortunately, Kongsted \textit{et. al.} did not study these higher excited states, which would have helped to determinate which results between our SCI-MDFT and EOM-CCSD/CSM are correct.

\begin{table}[h!]
\center
\begin{tabular}{c||c||cc}
                              & \multicolumn{1}{c||}{\textbf{QP}}          & \multicolumn{2}{c}{\textbf{EOM-CCSD}}       \\ \hline \hline
\multicolumn{1}{c||}{States}  &\multicolumn{1}{c||}{ \textbf{SCI}}  & \multicolumn{1}{c|}{\textbf{Howard}} &  \multicolumn{1}{c}{\textbf{Kongsted}}  \\ \hline
\multicolumn{1}{c||}{$\bm{^1A_2}$} &\multicolumn{1}{c||}{ 4.03} & \multicolumn{1}{c|}{4.03} & \multicolumn{1}{c}{4.0} \\ \hline
\multicolumn{1}{c||}{$\bm{^1B_2}$} &\multicolumn{1}{c||}{ 7.03} & \multicolumn{1}{c|}{7.03} & \multicolumn{1}{c}{N/A} \\ \hline
\multicolumn{1}{c||}{$\bm{^1B_2}$} &\multicolumn{1}{c||}{ 8.02} & \multicolumn{1}{c|}{7.98} & \multicolumn{1}{c}{N/A}\\ \hline
\multicolumn{1}{c||}{$\bm{^1A_1}$} &\multicolumn{1}{c||}{ 8.04} & \multicolumn{1}{c|}{8.05} & \multicolumn{1}{c}{N/A} \\ 
\end{tabular}
\caption{Excitation energies $\Delta E_i$ (a.u.) of formaldehyde in vacuum computed at the SCI level using aug-cc-pVDZ basis set compared to the predictions of Howard  \textit{et. al.}. and Kongsted \textit{et. al.} using EOM-CCSSD}
\label{tab:form_exc_vac}
\end{table}

\begin{table}[h!]
\center
\begin{tabular}{c||c|c|c||c|c}
& \multicolumn{3}{c||}{\textbf{QP/MDFT}}  & \multicolumn{2}{c}{\textbf{Litt.}} \\ \hline \hline
\textbf{States} & $\bm{\rho_{\infty}}$ & $\bm{\rho_2}$ & $\bm{\rho_0}$& \textbf{CCSD/PCM} & \textbf{CCSD/MD}$\bm{^*}$ \\ \hline
$\bm{^1A_2}$ & 4.14   & 4.15   & 4.17   & 4.32    &  4.34 \\ \hline
$\bm{^1B_2}$ & 7.59   & 7.61   & 7.63   & 7.09    &  -    \\ \hline
$\bm{^1B_2}$ & 8.52   & 8.54   & 8.63   & 8.03    &  -    \\ \hline
$\bm{^1A_1}$ & 8.59   & 8.61   & 8.77   & 8.13    &  -    \\ 
\end{tabular}
\caption{Excitation energies (a.u.) of the formaldehyde in aqueous solution. All calculations are made with the aug-cc-pVDZ basis set, except the EOM-CCSSD calculation (*) in which the aug-cc-pVTZ basis set were used.}
\label{tab:form_exc_solv}
\end{table}

\section{Conclusions \label{sec:Conclusion}}
In this paper, we propose to couple molecular density functional theory to advanced wavefunction methods as a way to account for solvent effects while computing excited states properties. While this approach belongs to the more general family of QM/MM method, its originality lies in the description of the solvent through its molecular density. This differs from the conventional MD method in which the solvent is depicted by a set of explicit molecules whose configurations need to be sampled. In practice, the equilibrium solvent density can be computed through functional minimization and the associated three dimensional charge density is directly obtained through a convolution product. 
From the point of view of the quantum solute, the solvent charge density is simply a set of additional point charges, creating the corresponding Coulomb potential that is added in the one body term in the Hamiltonian. 
We do not consider any Pauli repulsion or dispersion forces coming from the MM part; therefore, the interaction between the quantum solute and the MM part is purely electrostatic. 
On the other hand, the electronic density modifies the classical Hamiltonian through the electrostatic potential it generates. 
 To avoid the classical solvent to overlap with the quantum solute,  Lennard-Jones sites are added on top of the solute nuclei. This classical model of dispersion and repulsion creates the solute's cavity in the solvent. 
The sensibility of the QM/MM predictions on the LJ parameters have already been pointed point out\cite{jeanmairet_tackling_2020,rackers_classical_2019} but we resort here to parameters from standard MD forcefield without further investigation. 
Another important physical aspect that is missing from our description is the polarizability of the solvent molecules, which refers to how the electron cloud is altered by the presence of the solute and other solvent molecules.
Our QM/MM approach is performed self-consistently until both the MM solvation free energy and the QM ground state energy  converge. 
Once the solvent charge in equilibrium with the ground state is obtained, the excited states properties are computed using converged SCI 
calculations. This ensures that the electronic structure is described with a nearly exact precision. 
Therefore, the quality of the excitation energies can be directly attributed to the quality of the equilibrium 
classical charge density  and of the QM/MM coupling.
From the computational perspective, several QM and MDFT calculations are required to obtain the equilibrium solvent charge density.  
We would like to highlight that the cost of an MDFT calculation is independent of the complexity of the solute and relatively inexpensive compared 
to the 
QM calculation.  Therefore, it is desirable to use a QM method that provides reasonable computational efficiency.

We began our investigation by studying the quantum water molecule in the classical water solvent.  
We studied the impact of the quality of the ground state QM method (\textit{i.e.} HF, CISD an SCI) 
on the converged classical solvent charge by comparing the corresponding SCI excitation energies. 
The predicted SCI excitation energies appear to depend very weakly on the method chosen to optimize the solvent charge. This means that using Hartree-Fock is sufficient to converge the solvent charge. While this assertion have to be further checked for more complex solutes, particularly for strongly correlated systems, it is very encouraging to keep the computational cost of the introduced methodology tractable. 
The excitation energies predicted using SCI in the presence of the solvent charge obtained with QM/MDFT were found in satisfactory agreement 
with previously reported high quality QM/MM calculations\cite{kongsted_dipole_2002}. 

We emphasize that unlike the conventional CSM, the MDFT framework provides a detailed picture of the solvent structure surrounding the solute. For instance, it captures the complex structure of water around the water solute, as illustrated by the tetrahedral shape of the solvent charge density in the first solvation shell, depicted in Fig. \ref{fig_h2o_dens}.

We continued our study by computing the excitation energies of formaldehyde as similar EOM-CCSD/MD calculations have been previously reported. 
However, the quality of the results is here more reserved. We underestimate the first excitation energy by roughly 0.2 eV, and it is difficult to assess the quality of the prediction for the three following excited states due to the lack of references. 

These results suggest that further developments are needed to refine our QM/MDFT model. We believe that there are two main limitations to the presented approach. 
Firstly, a framework should be developed to account for the polarizability of the solvent molecules. 
This could be achieved by making the classical functional dependant on an additional field, namely the electronic polarization field. 
This field would naturally influence the external part of the functional, which account for the solute-solvent interaction but also the excess part, which models the solvent-solvent interaction. 
Secondly, the Pauli-dispersion interaction should be included in a more ab-initio manner, both in the quantum part and in the classical part. 
Overall, the MDFT description of the solvent coupled with advanced wavefunction methods to describe the solute appears to be a promising method 
to study electronic properties in solution, as it results in a simple one-body operator that could be coupled easily to any QM method or packages.

\section{Acknowledgements}
The authors want to thank Luc Belloni for providing the SPC/E direct correlation functions. M. L. acknowledges financial support from
Institut de Science des Mat\'eriaux (iMAT) at Sorbonne
Universit\'e.

\end{document}